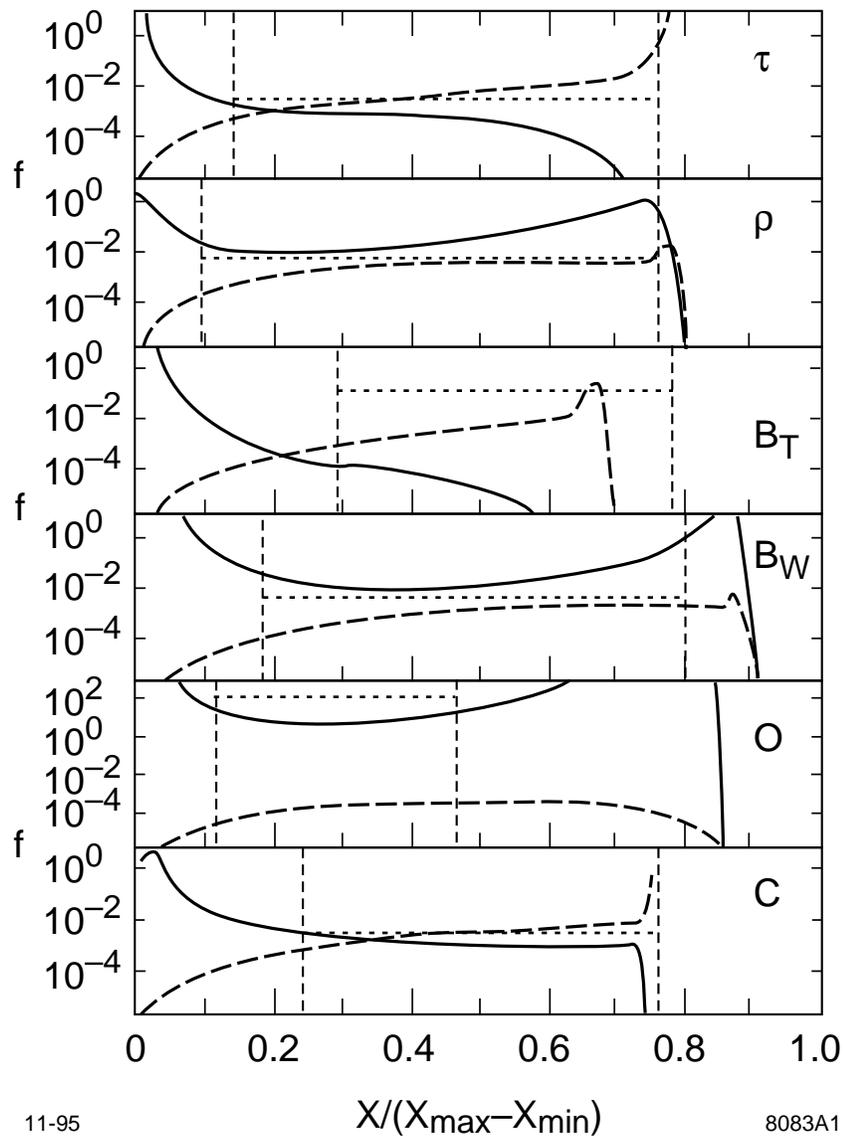

11-95  8083A1

Fig. 1a

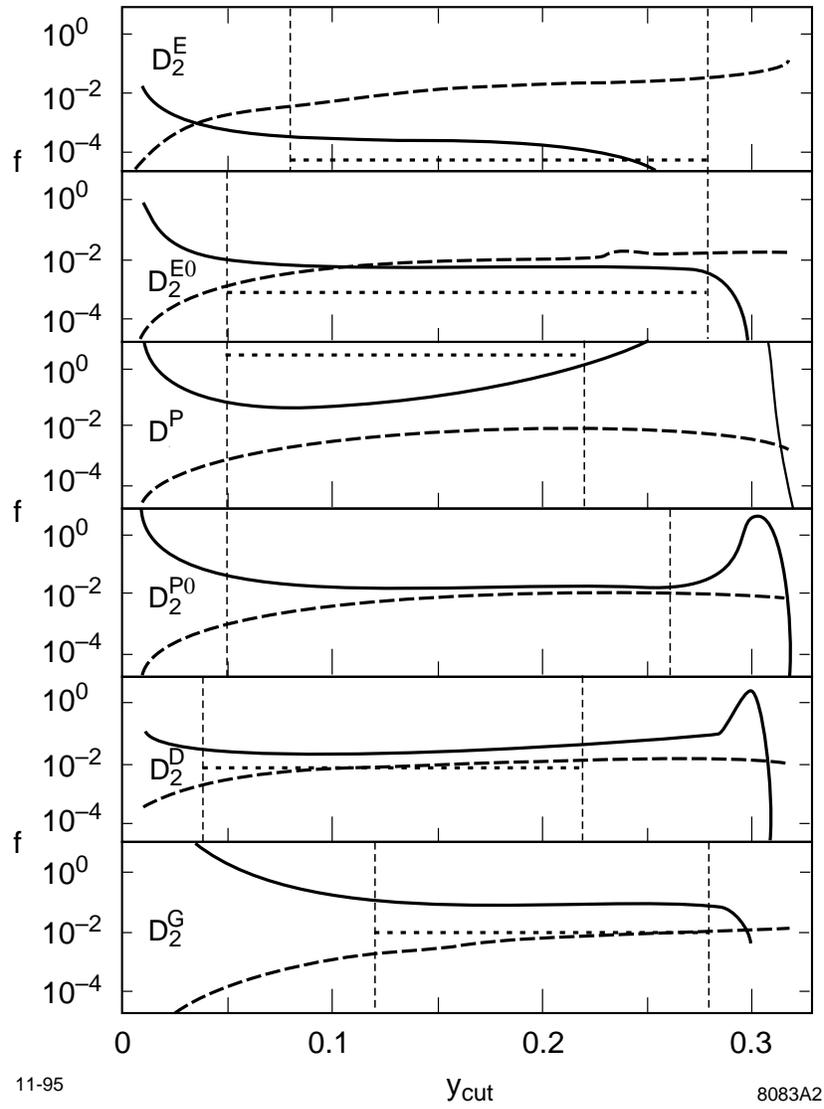

11-95  
8083A2

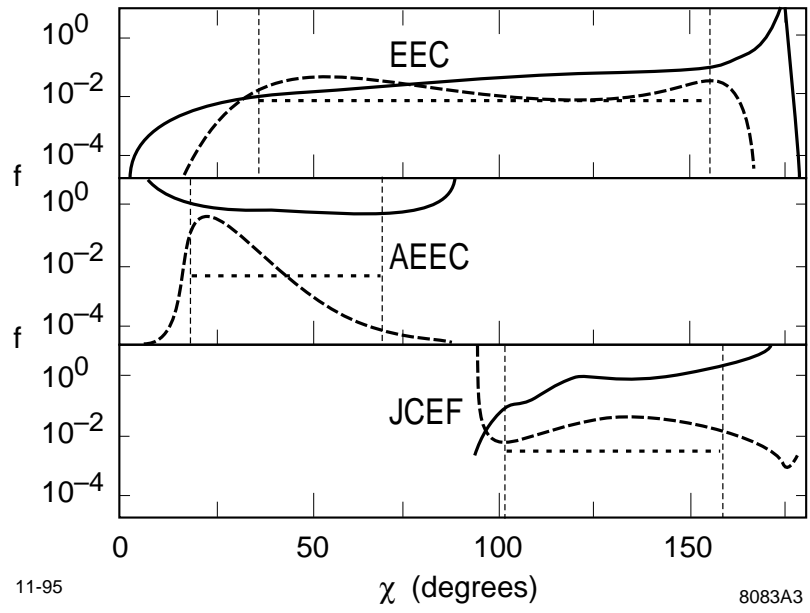

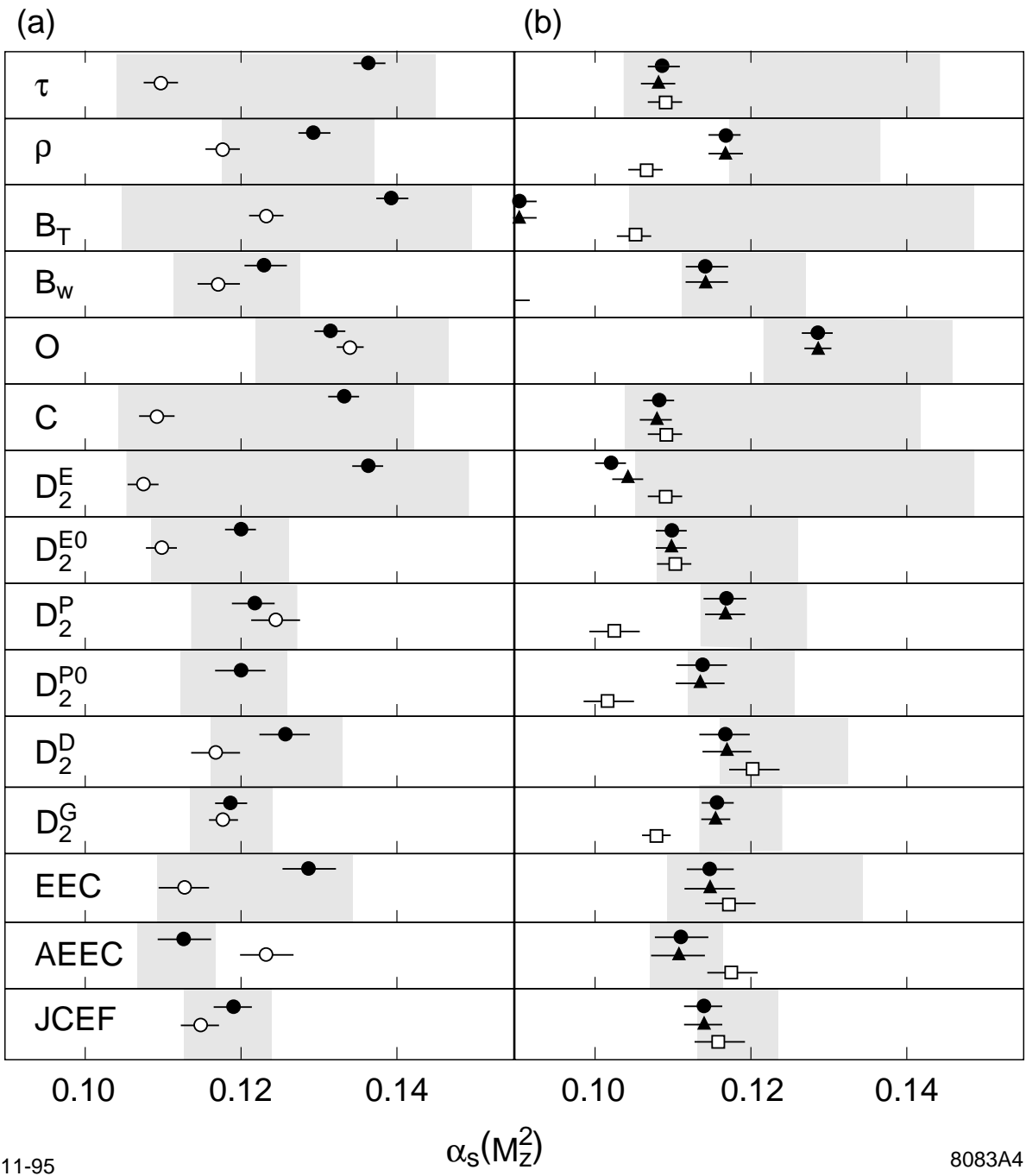

11-95

8083A4

# APPLICATION OF 'OPTIMISED' PERTURBATION THEORY TO DETERMINATION OF $\alpha_s(M_Z^2)$ FROM HADRONIC EVENT SHAPE OBSERVABLES IN e+e− ANNIHILATION[⋆]

P.N. Burrows
Massachusetts Institute of Technology
Cambridge, MA 02139, USA

H. Masuda, D. Muller
Stanford Linear Accelerator Center
Stanford University, Stanford, CA 94309, USA

Y. Ohnishi
Nagoya University
Nagoya 464, Japan

## ABSTRACT

We have applied so-called 'optimised' perturbation theory to resolve the renormalisation-scale ($\mu$) ambiguity of exact O($\alpha_s^2$) QCD calculations of event shape observables in e+e− → hadrons. We fitted the optimised predictions for 15 observables to hadronic $Z^0$ decay data from the SLD experiment to determine $\alpha_s(M_Z^2)$. Comparing with results using the physical scale $\mu = M_Z$ we found no reduction in the scatter among $\alpha_s(M_Z^2)$ values from the 15 observables, implying that the O($\alpha_s^2$) predictions with optimised scales are numerically no closer to the exact all-orders results than those with the physical scale.

*Submitted to Physics Letters B*

---

[⋆] Work supported by Department of Energy contracts DE–AC02–76ER03069 (MIT) and DE–AC03–76SF00515 (SLAC).

# 1. Introduction

The theory of strong interactions, Quantum Chromodynamics (QCD) [1], contains in principle only one free parameter, the strong interaction scale $\Lambda$. Tests of QCD hence comprise comparisons of measurements of $\Lambda$ in different processes and at different hard scales $Q$. In practise QCD calculations of observables are performed using finite-order perturbation theory, and calculations beyond leading order depend on the *renormalisation scheme* employed, implying a scheme-dependent $\Lambda$. Here we consider calculations performed in the modified minimal subtraction scheme ($\overline{MS}$ scheme) [2], and use the strong interaction scale $\Lambda_{\overline{MS}}$ for five active quark flavours. If one knows $\Lambda_{\overline{MS}}$ one may calculate the strong coupling $\alpha_s(Q^2)$ from the solution of the QCD renormalisation group equation [3]. Because of the large data samples taken in e$^+$e$^-$ annihilation at the $Z^0$ resonance, it has become conventional to use as a yardstick $\alpha_s(M_Z^2)$, where $M_Z$ is the mass of the $Z^0$ boson; $M_Z \approx 91.2$ GeV [4]. Tests of QCD can therefore be quantified in terms of the consistency of the values of $\alpha_s(M_Z^2)$ measured in different experiments.

Measurements of $\alpha_s(M_Z^2)$ have been performed in e$^+$e$^-$ annihilation, hadron-hadron collisions, and in deep-inelastic lepton-hadron scattering experiments, covering a range of $Q^2$ from roughly 1 to $10^4$ GeV$^2$; for recent reviews see [5,6]. Within the errors there is a remarkable degree of consistency between these measurements; an average yields $\alpha_s(M_Z^2) = 0.117 \pm 0.006$ [6], implying that QCD has been tested to a precision of about 5%. This precision is however rather modest compared with the achievement of sub-1% level tests of the electroweak theory [7], and this is due primarily to the *theoretical uncertainties* that dominate most of the experimental measurements. These uncertainties are due to both the restriction of complete perturbative QCD calculations to low order, and non-perturbative effects that are presently incalculable in QCD. The latter are often called 'hadronisation uncertainties' or 'higher twist effects'.

For any observable, truncation of the perturbation series at finite order causes a residual dependence on the (scheme-dependent) *renormalisation scale* $\mu$. In the case of e$^+$e$^-$ annihilation, perturbative QCD calculations of hadronic event shape observables, such as jet rates [8] or thrust [9], have been performed exactly only up to second order in $\alpha_s$ [10,11], and have been used extensively by experiments at the PETRA, PEP, TRISTAN, SLC, and LEP colliders for measurement of $\alpha_s(M_Z^2)$ [5,6]. The precision of these measurements is limited predominantly by the lack of knowledge of higher-order contributions, whose effect can, by definition, only be estimated in an *ad hoc* fashion. A consensus has arisen among experimentalists to estimate this effect from the $\mu$-dependence of the $\Lambda_{\overline{MS}}$, or equivalently $\alpha_s(M_Z^2)$, values derived from fits of the calculations to the data, see *eg.* [12], and to quote a corresponding *renormalisation scale uncertainty*. This procedure, well-motivated in that the $\mu$-dependence caused by the truncation of the perturbation series would be cancelled by addition of the higher-order terms, is, however, arbitrary, and is not equivalent to knowledge of the size of the *a priori* unknown terms. This arbitrariness is manifested in the fact that the experimental collaborations have chosen different ranges over which to vary $\mu$; combined with their different choices of observables and averaging methods, this has led to the variation among quoted central values of $\alpha_s(M_Z^2)$ and scale uncertainties shown in Table 1. Despite this variation it is apparent that the scale uncertainty is much larger than both the experimental error and the hadronisation uncertainty, and represents the most serious limitation towards improved precision on measurements of $\alpha_s(M_Z^2)$ using hadronic event shapes in e$^+$e$^-$ annihilation.

The renormalisation scheme/scale ambiguity of QCD calculations has been discussed extensively in the literature [17]. The scheme ambiguity can be sidestepped by simply adopting one particular scheme, usually $\overline{MS}$, as a reference; a next-to-leading order calculation of an observable in any scheme can then always be translated to the reference scheme [18]. The best resolution of the scale ambiguity would be to reduce its effect by calculating observables to higher order in perturbation theory. Though this is in principle possible, the large number of Feynman diagrams involved renders the task difficult and unattractive. In $e^+e^-$ annihilation only two inclusive observables, the hadronic cross-section ratio $R$ [19] and the $\tau$ hadronic decay ratio $R_\tau$ [20], have been calculated exactly up to $O(\alpha_s^3)$. For the hadronic event shape observables $O(\alpha_s^3)$ contributions have not yet been calculated completely, although progress has been made recently for some observables in the form of all-orders 'resummation' of leading and next-to-leading logarithmic contributions in the two-jet ('Sudakov') region [21]. These calculations have been 'matched' to the exact $O(\alpha_s^2)$ results to yield improved predictions which, though not formally complete at any order beyond the second, have been found to yield a reduced dependence on the renormalisation scale, but at the cost of additional uncertainties relating to ambiguities at $O(\alpha_s^3)$ in the matching procedure; see *eg.* [12]. In this paper we consider alternative approaches which attempt to 'optimise' $O(\alpha_s^2)$ QCD predictions by choosing the renormalisation scale according to *ad hoc* criteria.

## 2. Optimised Perturbation Theory

The $O(\alpha_s^2)$ prediction for an infra-red- and collinear-safe observable $X$ can be written:

$$\sigma_X \equiv \frac{1}{\sigma}\frac{d\sigma}{dX}(X,\mu) = \overline{\alpha_s}(\mu)A(X) + \overline{\alpha_s}^2(\mu)(A(X)2\pi b_0 \ln(\mu^2/Q^2) + B(X,N_f)) \quad (1)$$

where $\overline{\alpha_s} \equiv \alpha_s/2\pi$, $b_0 = (33 - 2N_f)/12\pi$, $N_f$ is the number of active quark flavours, $Q$ is the c.m. energy, and $A(X)$ and $B(X, N_f)$ can be calculated [11]. Explicit dependence on the renormalisation scale $\mu$ can be seen in the next-to-leading coefficient. We consider five possibilities for choosing this scale:

(i) *Physical scale:* $\mu = Q$, the hard scale of the interaction; in $e^+e^-$ annihilation at the $Z^0$ resonance this corresponds to $\mu = M_Z$. This choice explicitly removes the logarithmic term in Eq. (1).

(ii) *Experimentally-optimised scale:* $\mu$ can be derived from a simultaneous fit of $\Lambda_{\overline{MS}}$ and $\mu$ to experimental data. This is entirely pragmatic and allows the data to 'choose' the scale. The resulting $\Lambda_{\overline{MS}}$ and $\mu$ values are highly correlated [22].

(iii) *PMS scale:* Since the all-orders result would be independent of renormalisation scale, Stevenson suggests that $\mu$ be chosen according to the 'Principle of Minimal Sensitivity' (PMS) [23], from the solution of:

$$\frac{\partial \sigma_X}{\partial \mu} = 0. \quad (2)$$

(iv) *FAC scale:* Grunberg suggests that $\mu$ be chosen to give the 'fastest apparent convergence' (FAC) of the series [24], so that the second-order term in eq. (1) vanishes:

$$A(X) \, 2\pi b_0 \ln(\mu^2/Q^2) + B(X, N_f) = 0. \quad (3)$$

At next-to-leading order this is equivalent to the 'effective charge' (EC) approach [24,25].

(v) *BLM scale:* Brodsky, Lepage and Mackenzie advocate [26] that $\mu$ be chosen to remove the $N_f$-dependence of the second-order term in eq. (1):

$$\mu = Q \exp\{3(B(X, N_f = 5) - B(X, N_f = 4)/2A(X))\} \quad (4)$$

As in Quantum Electrodynamics, this effectively incorporates quark and gluon vacuum polarisation contributions into the definition of the strong coupling.

The experimentally-optimised, PMS, FAC and BLM approaches (ii)-(v) are usually collectively termed 'optimised' perturbation theory. In the PMS, FAC and BLM cases the optimised scale implicitly depends upon the value of the observable $X$.

Early theoretical studies applied optimised perturbation theory to jet rates in $e^+e^- \to$ hadrons, and employed the PMS approach at $Q \sim 34$ GeV [27], and the PMS, FAC and BLM approaches at $Q = M_Z$ [28]. Experimentally-optimised scales were also determined for jet rates at $Q = 29$ GeV [29]. The focus of these early studies was largely on obtaining an improved description of the rate of 4-jet production by $O(\alpha_s^2)$ QCD, which, for the physical scale $Q$, had been shown not to reproduce the PETRA data [8]. Until the SLC/LEP era, the influence of variation of the renormalisation scale on $\alpha_s(M_Z^2)$ measurements was usually not considered, corresponding, *de facto*, to choice of the physical scale. Early SLC/LEP $\alpha_s(M_Z^2)$ measurements based on jet rates and energy-energy correlations included experimentally-optimised scales [30] as well as the PMS, FAC and BLM methods [31,32]. However, theoretical controversy (see *eg.* [33]) motivated experimental groups to avoid specific scale-choice prescriptions, and subsequently to adopt the pragmatic approach of quoting an uncertainty on $\alpha_s(M_Z^2)$ by varying the renormalisation scale over a wide range; this approach has itself been criticised by theoreticians [34]. We are not aware of any comprehensive application of optimised perturbation theory to hadronic event shape observables, and of its effect on $\alpha_s(M_Z^2)$ determinations.

Here we present such a study. We have calculated the PMS, FAC and BLM scales for the 15 collinear- and infra-red-safe hadronic event shape observables considered in the SLD $\alpha_s(M_Z^2)$ measurement [12]. Using these scales, as well as the physical and experimentally-optimised scales, we have extracted $\alpha_s(M_Z^2)$ values from comparison with the SLD data. For each scale choice we have studied the scatter among the $\alpha_s(M_Z^2)$ values from the 15 observables, which one expects *a priori* to be reduced if the optimised perturbation series up to $O(\alpha_s^2)$ are indeed better approximations to the all-orders results.

## 3. Hadronic Event Shape Observables

First we review briefly the hadronic event shape observables. Thrust $T$ is defined [9]

$$T = \max \frac{\sum_i | \vec{p}_i \cdot \vec{n}_T |}{\sum_i | \vec{p}_i |}, \qquad (5)$$

where $\vec{p}_i$ is the momentum vector of particle $i$, and $\vec{n}_T$ is the thrust axis to be determined. We define $\tau \equiv 1 - T$. An axis $\vec{n}_{maj}$ can be found to maximize the momentum sum transverse to $\vec{n}_T$, and an axis $\vec{n}_{min}$ is defined to be perpendicular to the two axes $\vec{n}_T$ and $\vec{n}_{maj}$. The variables thrust-major $T_{maj}$ and thrust-minor $T_{min}$ are obtained by replacing $\vec{n}_T$ in Eq. (5) by $\vec{n}_{maj}$ or $\vec{n}_{min}$, respectively. The oblateness $O$ is then given by [35] $O = T_{maj} - T_{min}$. The $C$-parameter, $C = 3(\lambda_1\lambda_2 + \lambda_2\lambda_3 + \lambda_3\lambda_1)$, is derived from the eigenvalues $\lambda_i$ ($i$ = 1,2,3) of the infrared-safe momentum tensor [36]:

$$\theta_{\rho\sigma} = \frac{\sum_i p_i^\rho p_i^\sigma / | \vec{p}_i |}{\sum_i | \vec{p}_i |}, \qquad (6)$$

where $p_i^\rho$ is the $\rho$-th component of the three momentum of particle $i$, and $i$ runs over all the final state particles.

Events can be divided into two hemispheres, $a$ and $b$, of invariant mass $M_a$ and $M_b$, by a plane perpendicular to the thrust axis $\vec{n}_T$. The heavy jet mass $M_H$ is then defined [37] $M_H = \max(M_a, M_b)$. Here we consider the normalized quantity $\rho \equiv \frac{M_H^2}{E_{vis}^2}$, where $E_{vis}$ is the visible energy measured in each hadronic event. Jet broadening measures have been proposed in Ref. [38]. In each hemisphere $a$, $b$:

$$B_{a,b} = \frac{\sum_{i \in a,b} | \vec{p}_i \times \vec{n}_T |}{2 \sum_i | \vec{p}_i |} \tag{7}$$

is calculated. The total jet broadening $B_T$ and wide jet broadening $B_W$ are defined by $B_T = B_a + B_b$ and $B_W = \max(B_a, B_b)$, respectively.

For back-to-back two-parton final states $\tau$, $O$, $C$, $B_T$ and $B_W$ are zero; for planar three-parton final states $0 \leq \tau \leq 1/3$, $0 \leq O \leq 1/\sqrt{3}$ and $0 \leq C \leq 2/3$; spherical events have $\tau = 1/2$ and $C = 1$.

Another useful method of classifying the structure of hadronic final states is in terms of jets. Jets may be reconstructed using iterative clustering algorithms in which a measure $y_{ij}$, such as scaled invariant mass, is calculated for all pairs of particles $i$ and $j$, and the pair with the smallest $y_{ij}$ is combined into a single particle. This procedure is repeated until all pairs have $y_{ij}$ exceeding a value $y_{cut}$, and the jet multiplicity of the event is defined as the number of particles remaining. The $n$-jet rate $R_n(y_{cut})$ is the fraction of events classified as $n$-jet, and the differential 2-jet rate is defined [31]

$$D_2(y_{cut}) \equiv \frac{R_2(y_{cut}) - R_2(y_{cut} - \Delta y_{cut})}{\Delta y_{cut}}. \tag{8}$$

Several schemes have been proposed comprising different $y_{ij}$ definitions and recombination procedures. We have applied the E, E0, P, and P0 variations of the JADE algorithm [8] as well as the Durham (D) and Geneva (G) schemes [39].

Hadronic events can also be classified in terms of inclusive two-particle correlations. The energy-energy correlation ($EEC$) [40] is the normalized energy-weighted cross section defined in terms of the angle $\chi_{ij}$ between two particles $i$ and $j$ in an event:

$$EEC(\chi) \equiv \frac{1}{N_{events}\Delta\chi} \sum_{events} \int_{\chi-\frac{\Delta\chi}{2}}^{\chi+\frac{\Delta\chi}{2}} \sum_{ij} \frac{E_i E_j}{E_{vis}^2} \delta(\chi' - \chi_{ij}) \mathrm{d}\chi', \qquad (9)$$

where $\chi$ ($0 \leq \chi \leq 180°$) is an opening angle to be studied for the correlations, $\Delta\chi$ is the angular bin width, and $E_i$ and $E_j$ are the energies of particles $i$ and $j$ respectively. The shape of the $EEC$ in the central region, $\chi \sim 90°$, is determined by hard gluon emission. The asymmetry of the $EEC$ ($AEEC$) is defined as $AEEC(\chi) = EEC(180° - \chi) - EEC(\chi)$. Another procedure, related to the angle of particle emission, is to integrate the energy within a conical shell of opening angle $\chi$ about the thrust axis. The Jet Cone Energy Fraction ($JCEF$) is defined [41]:

$$JCEF(\chi) \equiv \frac{1}{N_{events}\Delta\chi} \sum_{events} \int_{\chi-\frac{\Delta\chi}{2}}^{\chi+\frac{\Delta\chi}{2}} \sum_{i} \frac{E_i}{E_{vis}} \delta(\chi' - \chi_i) \mathrm{d}\chi', \qquad (10)$$

where $\chi_i = \arccos{(\vec{p}_i \cdot \vec{n}_T / \mid \vec{p}_i \mid)}$ is the opening angle between a particle and the thrust axis vector, $\vec{n}_T$, whose direction is defined to point from the heavy jet mass hemisphere to the light jet mass hemisphere. Hard gluon emission contributes to the region corresponding to the heavy jet mass hemisphere, $90° \leq \chi \leq 180°$.

## 4. Measurement of $\alpha_s(M_Z^2)$

Distributions of these 15 event shape observables were measured [12] using a sample of approximately 50,000 hadronic $Z^0$ decay events collected by the SLD experiment. The data were corrected [12] for detector bias effects such as acceptance, resolution, and inefficiency, as well as for the effects of initial-state radiation and hadronisation, to arrive at 'parton-level' distributions.

For each observable we employed the EVENT program [42] to calculate the coefficients $A$ and $B$ in Eq. (1). We then fitted the $O(\alpha_s^2)$ calculation to the measured parton-level distributions, using first the physical scale $\mu = M_Z$, by minimising $\chi^2$ w.r.t. variation of $\Lambda_{\overline{MS}}$. In each case the fit range was chosen so as to exclude the 2-jet region, where resummation [21] of higher-order perturbative contributions is required [12], as well as the 4-jet region, where the $O(\alpha_s^2)$ calculation is not expected to reproduce the data accurately; these ranges are indicated in Fig. 1. Each resulting $\Lambda_{\overline{MS}}$ value was translated into $\alpha_s(M_Z^2)$; these are shown, with experimental errors [12], in Fig. 2(a). It can be seen that there is considerable scatter among the 15 $\alpha_s(M_Z^2)$ values. Since the same data sample was used to measure each observable, and since the observables are highly correlated, this scatter is very significant. Similar results have been observed previously [16]. The scatter can be interpreted as arising from uncalculated higher-order perturbative QCD contributions, which *a priori* may be of different sign and magnitude for the different observables. Taking an unweighted average over all 15 $\alpha_s(M_Z^2)$ values, and a corresponding r.m.s. deviation, yields[1]:

$$\alpha_s(M_Z^2) = 0.1265 \pm 0.0076 \quad \text{(physical scale)}.$$

We repeated this procedure using the experimentally-optimised-scale approach. In this case a simultaneous fit of $\Lambda_{\overline{MS}}$ and $\mu$ to each distribution was performed; the fitted value of $\mu$ is indicated in Fig. 1. The resulting pairs of $\Lambda_{\overline{MS}}$ and $\mu$ values were translated to $\alpha_s(M_Z^2)$, which are also shown in Fig. 2(a)[2]. Again, there is large scatter among the 15 $\alpha_s(M_Z^2)$ values. It should be noted that for most observables the experimentally-optimised scale yields a lower value of $\alpha_s(M_Z^2)$ than the physical scale; this is because the optimised scale is typically smaller than $M_Z$, which usually requires a smaller value of $\Lambda_{\overline{MS}}$ in order to fit the data [22]. For each observable such a systematic difference is encompassed by the renormalisation scale uncertainty on $\alpha_s(M_Z^2)$ considered in [12], which is also shown in Fig. 2. Taking an unweighted average and r.m.s. deviation yields[1]:

$$\alpha_s(M_Z^2) \quad = \quad 0.1173 \pm 0.0071 \qquad (\text{experimentally} - \text{optimised scale}).$$

---

[1] A weighted average based on experimental errors yields $\alpha_s(M_Z^2) = 0.1273$, which agrees with the unweighted average by less than the statistical error on $\alpha_s(M_Z^2)$ from a single observable.

[2] For the $D_2^{P0}$ observable no minimum in $\chi^2$ w.r.t. variation of $\mu$ in the range $10^{-4} \leq \mu^2/M_Z^2 \leq 10^2$ could be found.

As expected from the preceeding discussion, the central value is lower than for the physical scale choice. However, the r.m.s. deviation obtained with experimentally-optimised scales is comparable with that resulting from choice of the physical scale.

For each observable we then calculated the PMS, FAC and BLM optimised scales by solving Eqs. (2), (3) and (4), respectively; in the BLM case the next-to-leading coefficients $B$ were calculated separately using the EVENT program for $N_f = 4$ and 5. These scales are shown in Fig. 1. For each observable the following points are apparent: 1) the optimised scale depends strongly on the value of the observable; 2) with the exception of $O$, across the observable range the optimised scales are typically much smaller than the physical scale $\mu = M_Z$; 3) the PMS and FAC scales are almost identical; 4) as one approaches the 2-jet region, corresponding to $X/(X_{max} - X_{min}) \to 0$ in Fig. 1(a), $y_{cut} \to 0$ in Fig. 1(b), and $\chi \to 180°$ in Fig. 1(c), the BLM scale decreases whereas the PMS and FAC scales increase. In the last case the BLM behaviour conforms to the naive expectation that the optimised scale should be closely related to the momentum-transfer involved in the physical process, namely the radiation of soft and/or collinear gluons; the behaviour of the PMS and FAC scales in the two-jet limit does not appear to satisfy this expectation.

---

[1] A weighted average using experimental errors yields $\alpha_s(M_Z^2) = 0.1166$.

For each observable we then fitted the O($\alpha_s^2$) calculation to the measured distribution, using in turn the PMS, FAC and BLM scales, to determine $\Lambda_{\overline{MS}}$ and hence $\alpha_s(M_Z^2)$. The results are shown in Fig. 2(b); in the case of oblateness an acceptable fit with the BLM scale could not be obtained. For each observable it can be seen that the PMS- and FAC-derived $\alpha_s(M_Z^2)$ values are very similar, whereas, typically, the BLM-derived $\alpha_s(M_Z^2)$ value differs from them. This behaviour follows from the correlation between the scale value (Fig. 1) and the corresponding $\Lambda_{\overline{MS}}$ required to fit the data [22]. Comparing Figs. 2(a) and 2(b) it can be seen that for a given observable the PMS- and FAC-derived $\alpha_s(M_Z^2)$ values are often, though not always, close to that determined using the experimentally-optimised scale. Furthermore, for most observables the PMS-, FAC- and BLM-derived $\alpha_s(M_Z^2)$ values all lie within the range encompassed by the $\mu$-variation considered in [12], though for $\rho$, $B_W$, $D_2^P$, $D_2^{P0}$, $D_2^G$ and $(B_T)$, the BLM- (PMS/FAC-) derived values lie below this range.

The most striking feature of Fig. 2(b) is that, for any of the PMS, FAC or BLM scale choices, there is considerable scatter among the $\alpha_s(M_Z^2)$ values from all the observables. In each case, taking an unweighted average over all the $\alpha_s(M_Z^2)$ values and a corresponding r.m.s. deviation yields[1]:

$$\alpha_s(M_Z^2) \quad = \quad 0.1123 \pm 0.0079 \qquad \text{(PMS scale)}$$

---

[1] Weighted averages based on experimental errors yield central $\alpha_s(M_Z^2)$ values of 0.1120 (PMS), 0.1120 (FAC) and 0.1086 (BLM).

$$\alpha_s(M_Z^2) \quad = \quad 0.1123 \pm 0.0080 \qquad \text{(FAC scale)}$$

$$\alpha_s(M_Z^2) \quad = \quad 0.1088 \pm 0.0075 \qquad \text{(BLM scale)}.$$

In each case the r.m.s. deviation is comparable with that resulting from choice of the physical scale, or of the experimentally-optimised scale.

## 5. Summary and Conclusions

We have determined $\alpha_s(M_Z^2)$ by fitting $O(\alpha_s^2)$ QCD predictions of 15 hadronic event shape observables to $e^+e^-$ annihilation data at the $Z^0$ resonance collected by the SLD experiment. We used five prescriptions for resolving the renormalisation scale ambiguity that arises in the truncated perturbative calculation, namely the physical, experimentally-optimised, PMS-, FAC- and BLM-optimised scales. The average $\alpha_s(M_Z^2)$ value, taken over all the observables, differs among these five procedures, which can be understood from the correlation of $\alpha_s(M_Z^2)$ with the renormalisation scale value [22]. More importantly, the scatter among the $\alpha_s(M_Z^2)$ values is equally large for all five prescriptions, the r.m.s. deviation being about 0.008.

We conclude that the optimised perturbation theory approach does not reduce the scatter among the $\alpha_s(M_Z^2)$ values determined from different observables. If such scatter is interpreted as arising from the effects of the uncalculated higher-order perturbative QCD contributions, then in the cases we have investigated these contributions appear to be as large for optimised scales as for the physical scale. Therefore, notwithstanding the possible merits of optimised perturbation theory from a theoretical point-of-view, we have demonstrated that this approach does not appear to offer any numerical advantage in terms of the accuracy of perturbative QCD predictions of $e^+e^-$ event shapes. This is in agreement with the expectations of a recent study by Barclay and Maxwell [18], who advocate the use of renormalisation-scheme-invariant quantities as probes of the size of uncalculated higher-order QCD effects.

We thank our colleagues in the SLD Collaboration for support for this analysis. We also thank S. Brodsky, L. Dixon and C. Maxwell for helpful discussions and for their encouragement of this work.

| Experiment | Observables | $\alpha_s(M_Z^2)$ | Errors | | | Reference |
| | | | Exp. | Had. | Scale | |
| --- | --- | --- | --- | --- | --- | --- |
| SLD | $T$, $O$, $C$, $M_h^2$, $B_T$, $B_W$, $D_2^E$, $D_2^{E0}$, $D_2^P$, $D_2^{P0}$, $D_2^D$, $D_2^G$, EEC, AEEC, JCEF | 0.123 | ±0.003 | ±0.002 | ±0.011 | [12] |
| ALEPH | $D_2^{E0}$ | 0.121 | ±0.004 | ±0.007 a) | $^{+0.007}_{-0.012}$ | [13] |
| DELPHI | $T$, $O$, $C$, $M_h^2$, $M_d^2$, $D_2^{E0}$, EEC, AEEC | 0.113 | ±0.002 | ±0.003 | ±0.006 | [14] |
| L3 b) | $R_3^E$, $R_3^{E0}$, EEC, AEEC | 0.118 | ±0.004 | ±0.004 | ±0.006 | [15] |
| OPAL | $T$, $O$, $C$, $M_h$, $M_d$, $M_h^{(M)}$, $M_d^{(M)}$, $D_2^E$, $D_2^{E0}$, $D_2^P$, $D_2^D$, AEEC, PTEC | 0.122 | ±0.002 c) | – | $^{+0.006}_{-0.005}$ d) | [16] |

Table 1. $\alpha_s(M_Z^2)$ and errors from O($\alpha_s^2$) QCD fits to hadronic event shape observables in $Z^0$ decays. For each experiment results are taken from the most recent $\alpha_s$ determination based on event shapes using O($\alpha_s^2$) calculations. Definitions of the observables listed in the second column can be found in the references shown in the last column. The fourth column shows experimental errors, the fifth column hadronisation uncertainties and the sixth column scale uncertainties. Notes: a) uncertainty due to combined 'higher orders and hadronisation effects' [13], not including the scale uncertainty; b) we averaged the separate L3 measurements from jet rates and the EEC and AEEC; c) we estimated this value from Table 4 of Ref. [16]; d) Ref. [16] quotes a total uncertainty of $^{+0.006}_{-0.005}$ based on a weighted average over all 13 observables, taking correlations into account; subtracting the estimated experimental error in quadrature yields a theoretical uncertainty of $^{+0.006}_{-0.005}$, which includes both hadronisation and scale uncertainties.

**Figure Captions**

FIG. 1. Optimised values of the renormalisation scale plotted as $f = \mu^2/s$. Experimentally-optimised scale (horizontal dotted line); PMS and FAC (solid), and BLM (dashed) scales. Differences between the PMS and FAC scales cannot be resolved in this figure. In (a) the scale is plotted vs. the dimensionless variable $X/(X_{max} - X_{min})$, where $X = \tau$, $\rho$, $B_T$, $B_W$, $O$ or $C$, and $X_{max}$ ($X_{min}$) is the maximum (minimum) kinematically-allowed value of $X$. In (b) the scale is plotted vs. $y_{cut}$ for $D_2(y_c)$ calculated with the E, E0, P, P0, D and G jet-finding schemes. In (c) the scale is plotted vs. $\chi$ for the $EEC$, $AEEC$ and $JCEF$ (see text). The range of each observable used in the fits to determine $\alpha_s(M_Z^2)$ is indicated by vertical dashed lines.

FIG. 2. Values of $\alpha_s(M_Z^2)$ from QCD fits to the data using: (a) physical (solid circles), and experimentally-optimised (open circles) scales; (b) PMS- (solid circles), FAC- (solid triangles), and BLM- (open squares) optimised scales. In all cases only experimental error bars are shown. For each observable the shaded region indicates the total uncertainty estimated in Ref. [12], dominated by the contribution from wide variation of the renormalisation scale.